\documentclass{osa-article}

\usepackage{graphicx}
\usepackage{dsfont}
\usepackage{marvosym}
\usepackage{color}

\journal{osajournal}
\articletype{Research Article}

\newcommand{\ket}[1]{\left\vert{#1}\right\rangle}

\newcommand{\be}{\begin{equation}}
\newcommand{\ee}{\end{equation}}
\newcommand{\ba}{\begin{array}}
\newcommand{\ea}{\end{array}}
\newcommand{\bqa}{\begin{eqnarray}}
\newcommand{\eqa}{\end{eqnarray}}

\begin{document}

\title{Verifying genuine multipartite entanglement of the whole from its separable parts}

\author{Michal Mi\v{c}uda,\authormark{1,*} Robert St\'{a}rek,\authormark{1} Jan Provazn\'{\i}k,\authormark{1} Olga Leskovjanov\'{a},\authormark{1} and Ladislav Mi\v{s}ta, Jr.\authormark{1}}

\address{\authormark{1}Department of Optics, Palack\'{y} University, 17. listopadu 1192/12,  771~46 Olomouc, Czech Republic}

\email{\authormark{*}micuda@optics.upol.cz} 


\begin{abstract}
We prove experimentally the predicted existence of a three-qubit quantum state with genuine multipartite entanglement which can be certified solely from its separable two-qubit reduced density matrices. The qubits are encoded into different degrees of freedom of a pair of correlated photons and the state is prepared by letting the photons to propagate through a linear optical circuit. The presence of genuine multipartite entanglement is verified by finding numerically a fully decomposable entanglement witness acting nontrivially only on the reductions of the global state. Our result confirms viability of detection of emerging global properties of composite quantum systems from their parts which lack the properties.
\end{abstract}


\section{Introduction}
In many areas of human activity we are forced to infer a property of the whole from its fragments. On the macroscopic level the complexity of the investigated object together with poor partial information which we typically have in hands often precludes us from making an unambiguous statement about a global property of the object. The task is obviously easier provided that the parts which
we have at our disposal contain some marks of the analyzed property, but it is hardly tractable when they do not possess any evidences of the property.

A different situation occurs in the realm of quantum mechanics. Here, a relatively simple structure of composite quantum systems combined with utilization of rigorous mathematical tools frequently allow us to deduce from the reduced density matrices (marginals) whether a certain property is present in the global state, if it is absent, or it cannot be decided. More importantly, there are instances when one can prove the presence of a property in the global state even though the marginals do not carry any traces of the property. This is a particularly interesting scenario since such the property then can be viewed as an ``emergent'' property which arises from parts lacking the property \cite{Anderson_72}.

The purpose of the present paper is to investigate experimentally this astonishing feature of quantum world. We do it on a particularly important class of correlation properties in multipartite quantum systems known as multipartite entanglement. The concept of multipartite entanglement is a synonym for correlations among more than two systems, which cannot be prepared by local operations and classical communication and they stay behind, e.g.,non-statistical tests of quantum nonlocality \cite{Greenberger_89}, complex behaviour of strongly correlated systems \cite{Amico_08}, enhancement of precision of quantum measurements \cite{Giovannetti_04} or measurement-based model of quantum computing \cite{Raussendorf_01}. Applications such as quantum computing typically require an entangled state of a large number of quantum systems. However, with increasing number of participating systems it rapidly ceases to be feasible to reconstruct the global density matrix and one is sentenced to
infer global multipartite entanglement merely from more accessible marginals. The situation is easy if some marginal is entangled because then the entanglement is inherited also by the global state.
A much more interesting but also much more involved case arises when all marginals are separable. This is because then it is generally impossible to deduce unequivocally the presence or absence of entanglement in the global state as can be exemplified by the entangled three-qubit GHZ state $(|000\rangle+|111\rangle)/\sqrt{2}$ and by the fully separable state $(|000\rangle\langle000|+|111\rangle\langle111|)/2$, which both possess the same separable marginals $(|00\rangle\langle00|+|11\rangle\langle11|)/2$. Nonetheless, contrary to intuition,
sets of separable marginals can be constructed, which are compatible {\it only} with multipartite entangled global states \cite{Toth_05,Toth_07,Toth_09}.

The latter studies focused on certification of the presence of generic multipartite entanglement in the global state from separable marginals without any reference to the particular structure of the entanglement. However, practically all applications mentioned above are based on a special form of multipartite entanglement known as genuine multipartite entanglement (GME). For three qubits denoted as $A,B$ and $C$, which is a system considered in the rest of the present paper, the genuine multipartite entangled state is defined as a state which is not biseparable, i.e., it cannot be expressed as
\begin{eqnarray}\label{rhobisep}
\rho^{\rm bisep}_{ABC} &=& p_{1}\rho^{\rm sep}_{A|BC}+p_{2}\rho^{\rm sep}_{\rm B|AC}+p_{3}\rho^{\rm sep}_{C|AB},
\end{eqnarray}
where the state $\rho^{\rm sep}_{i|jk}=\sum_{n}p_{n}\rho_{i}^{n}\otimes\rho_{jk}^{n}$ is separable with respect to $i|jk$ bipartition and $p_{j}$, $j=1,2,3$, are probabilities. In Ref.~\cite{LChen_14} an example of a three-qubit state possessing all two-qubit marginals separable has been found using analytical tools, whose GME can be verified just from all the marginals. Only recently, a more systematic way of finding states with the sought property has been developed, which is based on the numerical finding of the GME witness acting only on two-qubit marginals of the state \cite{Miklin_16}. This resulted in the discovery of another three-qubit genuine multipartite entangled state with the desired property, which exhibits a larger robustness against addition of white noise compared to the original state of Ref.~\cite{LChen_14}.

So far, the property that GME can be detected only from the set of all marginals which are all separable has been analyzed only theoretically. In this paper we demonstrate the phenomenon experimentally by preparing the three-qubit quantum state of Ref.~\cite{Miklin_16}. The state is prepared using polarization and path degrees of freedom of a pair of correlated photons and probabilistic linear optical circuit. The separability of all two-qubit marginals of the prepared density matrix is verified by the partial transposition criterion \cite{Peres_96,Horodecki_96}, whereas to certify the presence of the GME we find numerically a fully decomposable entanglement witness which acts only on the marginals. Our result gives evidence that global properties of composite quantum systems
can be inferred merely from parts carrying no signatures of the properties.

The paper is organized as follows. In Sec.~\ref{sec_Theory} we give a concise explanation of the theoretical background of our experiment. Section~\ref{sec_Experiment} contains description of our
experimental setup, while Sec.~\ref{sec_Methods} is dedicated to the methodology of preparation of the investigated state. The results of the experiment are summarized in Sec.~\ref{sec_Results}
and Sec.~\ref{sec_Conclusions} contains conclusions.

\section{Theory}\label{sec_Theory}

The goal of the present paper is to demonstrate experimentally the possibility to detect GME in
a three-qubit quantum state from its separable two-qubit marginals. For this purpose we chose to prepare
the state of Ref.~\cite{Miklin_16}, which reads explicitly as:
\begin{eqnarray}\label{rho}
\rho &=& \frac{2}{3}|\xi\rangle\langle\xi|+\frac{1}{3}|\bar{W}\rangle\langle\bar{W}|,
\end{eqnarray}
where
\begin{eqnarray}\label{xi}
|\xi\rangle &=& \frac{1}{\sqrt{3}}|W^{\text{\tiny\Radioactivity}}\rangle+\sqrt{\frac{2}{3}}|111\rangle
\end{eqnarray}
with
\begin{eqnarray}\label{RadioactiveW}
|W^{\text{\tiny\Radioactivity}}\rangle &=& \frac{1}{\sqrt{3}}(e^{i\frac{\pi}{3}}|001\rangle+e^{-i\frac{\pi}{3}}|010\rangle-|100\rangle).
\end{eqnarray}
and
\begin{eqnarray}\label{barW}
|\bar{W}\rangle &=& \frac{1}{\sqrt{3}}(|011\rangle+|101\rangle+|110\rangle).
\end{eqnarray}

The first property that the state (\ref{rho}) has to possess is the separability of all two-qubit marginals $\rho_{jk}$, $jk=AB,BC,AC$. This property can be verified by the partial transposition criterion \cite{Peres_96,Horodecki_96}, according to which a two-qubit density matrix is separable if and only if its partial transposition is positive-semidefinite. Since all partial transposes $\rho_{jk}^{T_{j}}$ have the same positive lowest eigenvalues of  $\beta^{(jk)}:=\mathrm{min}[\mathrm{eig}(\rho_{jk}^{T_{j}})]=(5-2\sqrt{6})/27\doteq0.37\cdot10^{-2}$, all the marginals are separable as we set out to prove.

The second property is the presence of the GME in the state (\ref{rho}), which is verifiable from its separable marginals. This property can be certified by finding the so called GME witness which, in addition, acts nontrivially only on two-qubit marginals of the state \cite{Miklin_16}. For two parties an entanglement witness is an observable $\mathcal{W}$ with a non-negative average for all separable states and a negative average on at least one entangled state \cite{Horodecki_96,Terhal_00}. Here, we seek a witness of GME, which possesses a non-negative average on all biseparable states (\ref{rhobisep}) and there is an entangled state for which the average is negative. However, construction of the GME witness is in general a formidable task. Nevertheless, the specific feature of the state (\ref{rho}) is that its GME can be detected by a special kind of a witness called a fully decomposable entanglement witness \cite{Lewenstein_00,Jungnitsch_11} defined as a witness which can be for every subset $M$ of the set all three qubits $A,B$ and $C$ expressed as $\mathcal{W}=P_{M}+Q_{M}^{T_{M}}$, where $P_{M}$ and $Q_{M}$ are positive-semidefinite operators and $T_{M}$ is the partial transpose with respect to subsystem $M$. A fully decomposable entanglement witness is non-negative on the set of convex mixtures of states with positive partial transposition with respect to different bipartitions, which is a superset of the set of biseparable states and thus, states detected by the witness are genuinely multipartite entangled. A great advantage of fully decomposable witnesses is that they can be found numerically by solving a respective semidefinite programme (SDP) \cite{Vandenbergehe_96}. Together with the constraint on the structure of the sought witness the optimal witness with the desired properties can be found for the state (\ref{rho}) by solving the following SDP \cite{Miklin_16}:

\begin{equation}\label{SDP}
\begin{aligned}
& \underset{\mathcal{W},P_{M},Q_{M}}{\text{minimize}}
& & \mathrm{Tr}(\rho \mathcal{W}) \\
& \text{subject to}
& &\mathrm{Tr}(\mathcal{W})=1,\,\, \mbox{where}\\
& & & \mathcal{W}=\sum_{i,j=0}^{3}w_{ij}^{(AB)}\sigma_{i}^{(A)}\otimes\sigma_{j}^{(B)}\otimes\mathds{1}^{(C)}+\mbox{permutations},\\
& & & \mbox{and for all bipartitions $M|\overline{M}$},\\
& & & \mathcal{W}=P_{M}+Q_{M}^{T_{M}},\quad P_{M}\geq0,\quad Q_{M}\geq0.
\end{aligned}
\end{equation}

The first constraint is just the normalization condition on the witness. In the second constraint the symbol $\sigma_{k}^{(l)}$ stands for the Pauli-$k$ matrix of a qubit $l$ and the constraint guarantees that the witness acts only on two-qubit marginals of the state (\ref{rho}). Finally,  the third constraint ensures full decomposability of the witness. We solved the SDP (\ref{SDP}) numerically using the YALMIP \cite{yalmip} package to interface with the MOSEK \cite{mosek} solver (obtained under personal academic licensing option). We adapted the PPTMixer \cite{PPTMIXER} to support a recent version of YALMIP and introduced the second constraint from (\ref{SDP}). In particular, for the state (\ref{rho}) we get explicitly $\mathrm{Tr}(\rho \mathcal{W})\doteq-1.98\cdot10^{-2}$, which confirms the presence of GME verifiable from separable marginals.

Before moving to the experiment needless to say, that due to the closeness of the eigenvalues $\beta^{(jk)}$ to zero, the reduced states $\rho_{jk}$ lie very close to the set of entangled states. For the sake of unambiguous demonstration of the desired effect, we add a certain fraction of the white noise to the density matrix (\ref{rho}), thus obtaining
\begin{equation}\label{rhop}
\rho_p=p\rho+\frac{(1-p)}{8}\mathds{1},\quad 0\leq p\leq 1,
\end{equation}
where $\mathds{1}$ is the $8\times8$ identity matrix. Naturally, by decreasing the mixing parameter $p$ the eigenvalue $\beta^{(jk)}$ increases but at the same time the mean value of the obtained entanglement witness also increases. However, as we have already said before, the density matrix (\ref{rho}) exhibits white-noise tolerance of up to 13.7\% and thus the investigated effect survives addition of even a moderate amount of white noise \cite{Miklin_16}. Below in the experimental part of the present paper we optimize the value of the parameter $p$ so as to observe both the needed properties with equally high statistical significance.

\section{Experiment}\label{sec_Experiment}

\begin{figure}[ht]
\begin{center}
\includegraphics[width=0.99\textwidth]{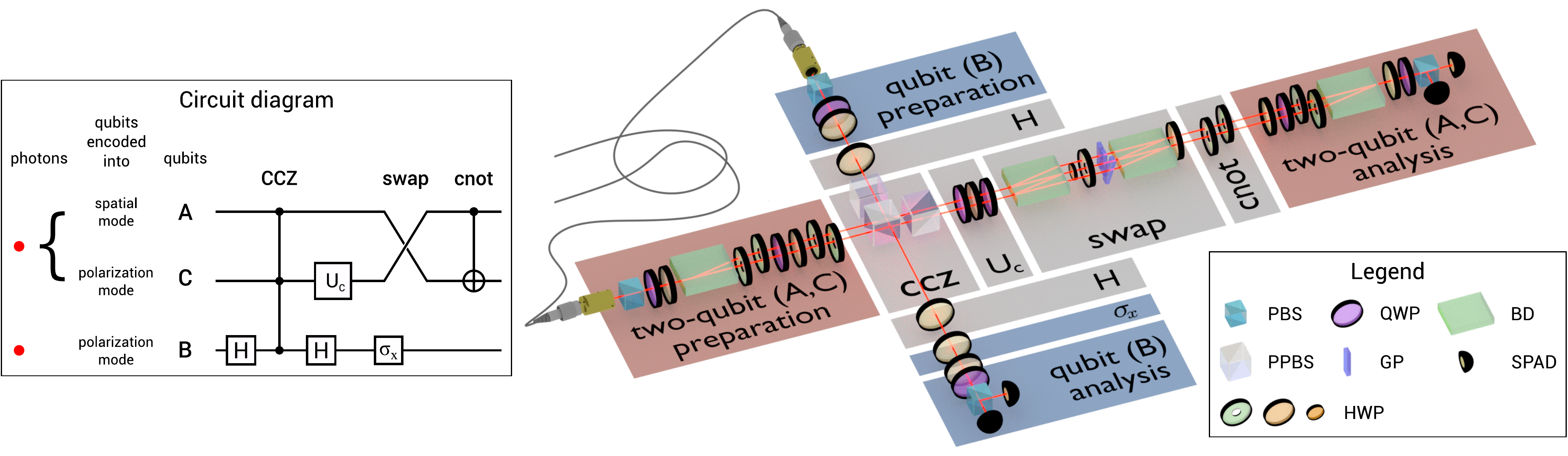}
\caption{Symbolic circuit diagram and experimental scheme of photonic circuit producing three-qubit quantum state $\rho_p$, Eq.~(\ref{rhop}). Red line denotes photon's path. HWP -- half-wave plate (the ring variant is used in cnot and swap blocks), QWP -- quarter-wave plate, BD -- calcite beam displacer, PPBS -- partially polarizing beam splitter, PBS -- polarizing beam splitter, SPAD -- collimator and single-mode optical fiber connected to single-photon avalanche photodiode. See text for details.}
\label{setup}
\end{center}
\end{figure}

The desired three-qubit quantum state $\rho$, Eq.~(\ref{rho}), is prepared by a photonic circuit depicted in Fig.~\ref{setup} based on quantum linear optics platform~\cite{Knill01}. We encoded three-qubits into two photons using hyper-encoding technique which exploits encoding of two qubits into polarization and path degrees of freedom of a single photon. We used orthogonally polarized time-correlated photon pairs generated in the process of spontaneous parametric down-conversion in a nonlinear crystal pumped by a cw laser diode. The first qubit $A$ is encoded into the spatial degree of freedom while the second qubit $C$ is encoded into the polarization degree of freedom of the single photon. The spatial qubit $A$ is initially prepared in polarization encoding using combination of quarter-wave plate (QWP) and half-wave plate (HWP), and it is subsequently converted into path encoding using a calcite beam displacer (BD). The computational state $\ket{0}$ corresponds to the horizontally polarized photon propagating in the upper interferometer arm, while the state $\ket{1}$ is represented by a vertically polarized photon propagating in the lower interferometer arm of an inherently stable MZ interferometer formed by the first and the last BD~\cite{OBrien03, Lanyon08, Lanyon10}. More details about the phase stability of the interferometric setup mentioned above is discussed in Ref.~\cite{Starek18}. The polarization qubit $C$ is prepared by combination of the next QWP and HWP placed into both arms of the MZ interferometer. The second photon served as a carrier for qubit $B$, which is encoded into the polarization state using combination of the QWP and HWP.

The photonic circuit is made up of four single and two two-qubit gates, and one three-qubit gate. Even though all single and two-qubit gates allow deterministic operation, the overall scheme is probabilistic due to the nature of the used three-qubit CCZ gate which is realized by two-photon interference on an unbalanced beam splitter~\cite{Ralph02, Kiesel05, Langford05, Okamoto05, Micuda13, Micuda15}. The CCZ gate is the only probabilistic gate in the setup and limits the success probability to the theoretical value of $1/9$. Every single- and two-qubit quantum gate can be independently switched on or off as required. This enabled us to sequentially prepare both parts of the state $\rho$, Eq.~(\ref{rho}). The Hadamard gates $H$ are realized by HWP rotated by $\pi/8$ radians, which changes computation basis $\{\ket{0}$, $\ket{1}\}$ to $\{\ket{+}\equiv(|0\rangle+|1\rangle)/\sqrt{2}$, $\ket{-}\equiv(|0\rangle-|1\rangle)/\sqrt{2}\}$. The unitary operation $U_C$ is implemented by a sequence of a QWP, HWP, and a QWP. The $\sigma_x$ operation is achieved by HWP rotated by $\pi/4$ radians. The two-qubit swap gate operates on qubits $A$ and $C$, which are hyper-encoded in the signal photon \cite{Starek18}. This operation interchanges the path and the polarization qubit and it is realized by sandwiching two HWPs rotated by $\pi/4$ radians between two calcite beam displacers, which creates two MZ interferometers with one common arm. Both HWPs were realized by a single ring-shaped HWP with a circular gap in the center placed in such the way that the common arm passes unaltered. The cnot gate between qubits $A$ and $C$, where $A$ is control, is implemented by one HWP rotated to $\pi/4$ radians and the second HWP is unaltered. This flips the state of the polarization qubit but only for a single spatial qubit state. This easy implementation is enabled by the hyper-encoding of qubits and offers unparalleled process fidelity \cite{Micuda15}.

The qubits $A$ and $C$ are analyzed by two-qubit analysis consisting of a HWP, QWP, BD, and the standard polarization analysis block consisting of a HWP, QWP, and polarizing beam splitter (PBS). The qubit $B$ is analyzed by the standard polarization analysis block. Output beams are coupled to single-mode optical fibers, which provide spatial filtering, and they are guided to the avalanche photo diode (SPAD) detectors. The electronic signal from the SPAD detectors is electronically processed through delay lines and coincidence logic and the number of coincidences is recorded by an electronic counter.

\section{Methods}\label{sec_Methods}

Owing to the closeness of separable two-qubit reductions to the set of entangled states, we experimentally prepared the mixture~(\ref{rhop}). To have full control over the structure of prepared states we have separately prepared the state $\ket{\xi}$, Eq.~(\ref{xi}), the state $\ket{\bar{W}}$, Eq.~(\ref{barW}), and all states $|000\rangle$, $|001\rangle$, ..., $|111\rangle$ of the computational basis representing diagonal elements of the identity matrix.

To generate the state $\ket{\xi}$ we prepared qubits $A$, $B$ and $C$ initially in the state
\bqa\label{xi_prepar}
\left[\frac{\sqrt{2}}{3} \ket{11}_{AC} + \frac{1}{3\sqrt{2}} ( e^{-i\frac{\pi}{3}} + \sqrt{6}e^{i\frac{\pi}{3}} ) \ket{00}_{AC}+\frac{1}{\sqrt{2}} ( e^{-i\frac{2\pi}{3}} - \sqrt{6} ) \ket{01}_{AC}\right]\otimes \ket{0}_{B}
\eqa
and then we propagated them through the logical circuit depicted in Fig.~\ref{setup}, where the unitary operation $U_C$ reads as

\be\label{operations}
 U_{C} = \frac{1}{\sqrt{2}}
 	\left(
 		\ba{cc}
		-1 & e^{-i\frac{\pi}{3}} \\
		e^{i\frac{\pi}{3}} & 1
 		\ea
 	\right).
\ee

The prepared state was characterized by three-qubit quantum state tomography which consists of sequential projections onto the six states $|0\rangle$, $|1\rangle$, $|+\rangle$, $|-\rangle$, ($|0\rangle + i|1\rangle)/\sqrt{2}$, ($|0\rangle - i|1\rangle)/\sqrt{2}$ at each output qubit for total $6^3=216$ measurements. Two-photon coincidences were recorded for $1\,\rm{s}$ for each measurement and yielded approximately $350$ pairs in average. The measured coincidence counts were normalized by sum of all coincidences and relative frequencies were obtained. The density matrix was reconstructed from this data using maximum likelihood estimation algorithm~\cite{Jezek03, Paris04}. The quality of the prepared state can be characterized by quantum state fidelity, given by

\be\label{statefid}
F = \mathrm{Tr}[(\rho^{1/2}\; \rho^{\rm exp} \rho^{1/2})^{1/2}]^2,
\ee

where $\rho$ is the ideal theoretical state and $\rho^{\rm exp}$ is the experimental state. One source of reduction in the quantum state fidelity is the introduction of phase shifts experienced by one or more modes in the setup, which are caused by the imperfect nature of the realistic experimental components. This can be compensated for by suitable unitary corrections. To reflect this, we introduce the optimized fidelity, which was calculated from the experimental state subject to three unitary operations, one in each of the input modes. The three unitary operations were optimized over and ultimately chosen in such a way that the resulting fidelity is maximal. The knowledge of the unitary operations enabled us to incorporate them into three-qubit preparation stage by physically changing the input state~(\ref{xi_prepar}) and thus improving quality of produced state $\ket{\xi}$. This procedure allowed to increase quantum state fidelity of prepared state $\ket{\xi}$ from $0.620$ to $0.935$. We also evaluated the purity of the state as $P = \mathrm{Tr}[(\rho^{\rm{exp}})^2]$, which was $0.93$. For estimation of uncertainty of the experimental results we have used standard Monte Carlo analysis. Using measured coincidence counts as a mean value of Poisson distribution we have numerically generated $10^4$ samples of the state, which were again reconstructed with the help of the Maximum Likelihood estimation algorithm. The fidelity and purity of the prepared state $\ket{\xi}$ were $0.930(8)$ and $0.92(1)$, respectively, including one standard deviation related to the last significant digit represented by the number in the brackets.

The state $\ket{\bar{W}}$, Eq.~(\ref{barW}), was prepared in a similar way. Qubits $A$, $B$ and $C$ were prepared initially in the state
\be\label{W_prepar}
\frac{1}{\sqrt{3}}\left(\ket{01}_{AC} + \sqrt{2}\ket{1}_A \ket{+}_C\right)\otimes \ket{0}_B
\ee
and subsequently propagated through the logical circuit displayed in Fig.~\ref{setup}, but with the operation $U_C$, swap gate and cnot gate
switched off. Using the method described above we were able to increase quantum state fidelity of prepared state $\ket{\bar{W}}$ from $0.858$ to $0.912$ with the purity of $ 0.92$. The Monte Carlo analysis yielded $F = 0.909(8)$ and $P = 0.91(1)$. Combining acquired relative frequencies for states $\ket{\xi}$ and $\ket{\bar{W}}$ we were able to prepare state~(\ref{rho}) with fidelity $0.89(1)$ and purity $0.473(9)$, while ideal purity is equal to $5/9$.

The diagonal elements of the identity matrix were measured on the full logical circuit shown in Fig.~\ref{setup} and they were not subjected to compensation by suitable unitary operations. The fidelity and purity of the identity were $ 0.9775$ and $0.1368$, respectively. Please note, that purity of the ideal identity is equal to $P = 0.125$. The Monte Carlo analysis yielded $F = 0.9774(4)$ and $P = 0.1369(2)$. Results obtained by the Monte Carlo analysis are in an excellent agreement with the measured experimental values for all three states generated by our optical circuit.

\section{Results}\label{sec_Results}

Successful demonstration of the investigated phenomenon requires to prove two properties of the prepared state.
First, all two-qubit marginals of the state have to be separable and second, it also has to contain GME which can
be detected solely from the marginals. In order to verify both the properties unambiguously and with highest attainable
statistical significance, we have prepared experimentally mixtures (\ref{rhop}) with the mixing parameter $p$ from the interval
containing the values for which the desired effect occurs. The experimental density matrices $\rho^{\mathrm{exp}}_{p}$ have been tomographically
reconstructed using the maximum likelihood estimation algorithm and the uncertainties in the quantities have been
estimated with the help of the Monte Carlo simulation. Separability of all two-qubit reductions of $\rho^{\mathrm{exp}}_{p}$ has been analyzed by
calculating the lowest eigenvalue of partial transposes of all three marginals, which shows up to be
always the eigenvalue $\mathrm{min}\{\mathrm{eig}[(\rho_{p,BC}^{\mathrm{exp}})^{T_{B}}]\}$.

\begin{figure}[!ht!]
	\centerline{\includegraphics[width=0.99\textwidth]{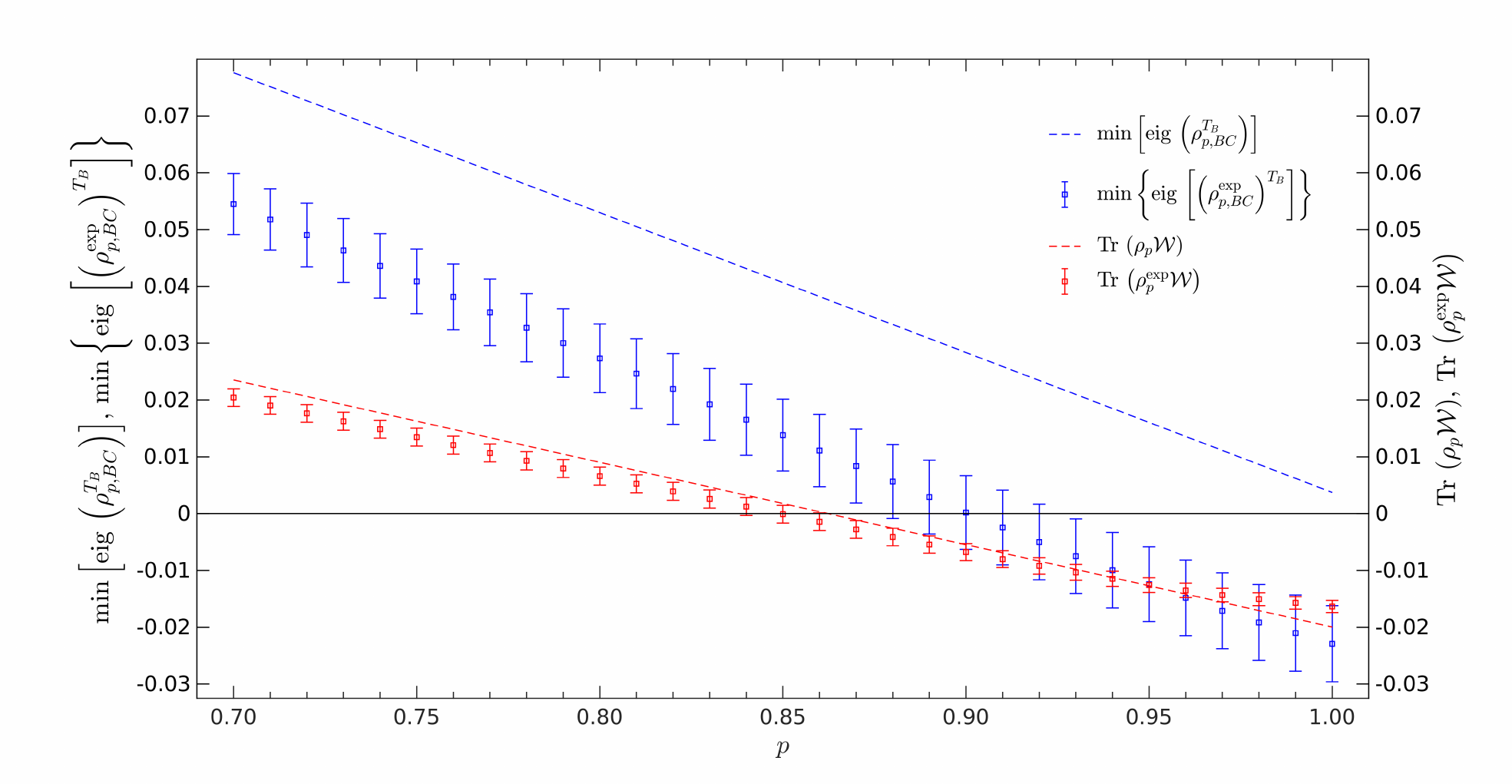}}
	\caption{Dependence of the theoretical lowest eigenvalue $\mathrm{min}[\mathrm{eig}(\rho_{p,BC}^{T_{B}})]$ (dashed blue line), the experimental lowest eigenvalue $\mathrm{min}\{\mathrm{eig}[(\rho_{p,BC}^{\mathrm{exp}})^{T_{B}}]\}$ (blue squares), the theoretical mean value $\mathrm{Tr}(\rho_{p} \mathcal{W})$ (dashed red line) for the state (\ref{rhop}) and the experimental mean value $\mathrm{Tr}(\rho^{\mathrm{exp}}_{p} \mathcal{W})$ (red squares) for the experimental state $\rho^{\rm exp}_{p}$ as a function of the mixing parameter $p$. See text for details.}
	\label{pplot}
\end{figure}

As for the presence of GME it is typically verified by a direct measurement of the witness operator. However, here we could not follow this route because
the explicit form of the witness is {\it a priori} not known. Instead, we numerically found the optimal entanglement witness $\mathcal{W}_{p}^{\mathrm{exp}}$ for each experimental density matrix $\rho_{p}^{\mathrm{exp}}$ by solving the SDP (\ref{SDP}). We then calculated the expectation value $\mathrm{Tr}(\rho_{p}^{\mathrm{mc}, i} \mathcal{W}_{p}^{\mathrm{exp}})$ for every density matrix $\rho_{p}^{\mathrm{mc}, i}$ of the $10^4$ matrices generated by the Monte Carlo simulation and used these values to determine the standard deviation for each mixing parameter $p$. The obtained results are summarized in Fig.~\ref{pplot}.

From the figure and the numerical calculation we identified the value of the mixing parameter $p=0.868$, for which the values of the quantities certifying the
required properties beat the bounds given by zero values roughly by $1.5$ standard deviations. The values of the lowest eigenvalues of the partial transposes of all three two-qubit marginals of the experimental state with $p=0.868$ are summarized in Table~\ref{table1}.

\begin{table}[ht]
\caption{Minimal eigenvalue $\beta_{\rm exp}^{(jk)}:=\mathrm{min}\{\mathrm{eig}[(\rho^{\mathrm{exp}}_{0.868,jk})^{T_{j}}]\}$
with one standard deviation of the partial transpose with respect to qubit $j$ of the reduced density matrix
$\rho^{\mathrm{exp}}_{0.868,jk}$ of qubits $j$ and $k$:} \centering
\begin{tabular}{| c | c | c | c |}
\hline $jk$ & AB & BC & AC \\
\hline $\beta_{\rm exp}^{(jk)}\cdot10^2$ & $2.35\pm0.70$ & $0.94\pm0.64$ & $2.12\pm0.65$ \\
\hline
\end{tabular}
\label{table1}
\end{table}
The Table~\ref{table1} reveals that all the partial transposes are strictly positive and thus all two-qubit marginals of the prepared state are separable as required.

Likewise, for the numerically found optimal fully decomposable entanglement witness $\mathcal{W}$ acting only on two-qubit reductions we get $\mathrm{Tr}(\rho^{\mathrm{exp}}_{0.868} \mathcal{W})=(-2.5\pm1.6)\cdot 10^{-3}$ which demonstrates the presence of the second property in the investigated state. The numerically calculated mean value of the witness for the experimentally prepared state is rather small but needless to say that the precision of the SDP solver is at least $10^{-6}$ \cite{mosek}.

Our experiment aimed at demonstration of a fragile phenomenon consisting of simultaneous
presence of two mutually opposing properties and therefore we succeeded to verify it
with a significance level of only $1.5\,\sigma$. The relevant quantities were calculated
using the density matrix obtained by the maximum likelihood reconstruction algorithm
which suffers from bias~\cite{Schwemmer15}. This introduces a systematic error into the quantities of our interest, which we analyze in Supplement 1. We performed numerical simulations which reveal that maximum-likelihood reconstruction algorithm shifts $\mathrm{Tr}(\rho\mathcal{W})$ towards positive values and all $\beta^{(jk)}$ towards negative values. To claim that the state $\rho$ has genuine multipartite entanglement verifiable from separable marginals, conditions $\mathrm{Tr}(\rho\mathcal{W}) < 0 $ and $\beta^{(jk)} \geq 0$ for all $jk$ have to be fulfilled simultaneously. In this context, it is plausible to consider both the quantities obtained by the maximum-likelihood method to be conservative estimates. In other words, the significance level with which we have demonstrated the property can be slightly underestimated. Besides, the simulations also suggest, that for the experimental count rates the systematic errors for all the discussed quantities are lower than one half of a standard deviation. Despite the systematic errors, nearly $79\%$ of the Monte Carlo samples satisfy simultaneously the conditions $\mathrm{Tr}(\rho\mathcal{W}) < 0 $ and $\beta^{(jk)} \geq 0$, for all $jk$.

\section{Conclusions and outlook}\label{sec_Conclusions}

We have demonstrated experimentally the possibility to verify GME of a three-qubit quantum state from its separable marginals. In our experiment, the certification of the presence of the GME relied on reconstruction of the global density matrix followed by numerical calculation of the GME witness acting on {\it all} two-qubit marginals. The question that naturally arises in this context is as to whether the phenomenon can be demonstrated also by a direct measurement of the respective witness operator. Another interesting open question concerns the possibility to detect experimentally GME by measuring not all but only some separable marginals as it was proposed theoretically in Ref.~\cite{Paraschiv_18}. A large and completely unexplored area of research eventually comprise questions
related to the existence and experimental observability of the investigated phenomenon in the realm of Gaussian states. We believe that our result will stimulate further studies on detection of global correlation as well as other properties of complex quantum systems from their parts which lack the properties.

\section{Funding}
This work was supported by the Czech Science Foundation (GA16-17314S). Robert St\'{a}rek and Jan Provazn{\'{i}}k acknowledge support from IGA-PrF-2018-010 and IGA-PrF-2019-010.

\end{document}